\documentclass[lettersize,journal]{IEEEtran}
\usepackage{amsmath,amsfonts}
\usepackage{algorithmic}
\usepackage{algorithm}
\usepackage{array}
\usepackage[caption=false,font=normalsize,labelfont=sf,textfont=sf]{subfig}
\usepackage{textcomp}
\usepackage{stfloats}
\usepackage{url}
\usepackage{verbatim}
\usepackage{graphicx}
\usepackage{xcolor}
\usepackage[english]{babel}
\usepackage[autostyle]{csquotes}
\usepackage{cite}
\usepackage{mathtools}

\newlength\myindent
\newfont{\bbb}{msbm10 scaled 500}

\newfont{\bb}{msbm10 scaled 1100}

\newcommand{\RR}{\mbox{\bb R}}

\newcommand{\pv}{{\bf p}}

\newcommand{\Bm}{{\bf B}}

\newcommand{\Dm}{{\bf D}}

\newcommand{\Id}{{\bf I}}

\newcommand{\Km}{{\bf K}}

\newcommand{\Mm}{{\bf M}}

\newcommand{\deltav}{\hbox{\boldmath$\delta$}}

\newcommand{\epsilonv}{\hbox{\boldmath$\epsilon$}}

\newcommand{\thetav}{\hbox{\boldmath$\theta$}}

\newcommand{\omegav}{\hbox{\boldmath$\omega$}}

\usepackage{multirow}
\def\BibTeX{{\rm B\kern-.05em{\sc i\kern-.025em b}\kern-.08em
    T\kern-.1667em\lower.7ex\hbox{E}\kern-.125emX}}
\usepackage{float}
\usepackage{caption}

\newcommand\algorithmicprocedure{\textbf{procedure}}
\newcommand{\algorithmicendprocedure}{\algorithmicend\ \algorithmicprocedure}
\makeatletter
\newcommand\PROCEDURE[3][default]{%
  \ALC@it
  \algorithmicprocedure\ \textsc{#2}(#3)%
  \ALC@com{#1}%
  \begin{ALC@prc}%
}
\newcommand\ENDPROCEDURE{%
  \end{ALC@prc}%
  \ifthenelse{\boolean{ALC@noend}}{}{%
    \ALC@it\algorithmicendprocedure
  }%
}
\newenvironment{ALC@prc}{\begin{ALC@g}}{\end{ALC@g}}
\makeatother

\begin{document}

\title{Localizing Load-Altering Attacks Against Power Grids Using Deep Capsule Nets}


\author{Hamidreza Jahangir, Subhash Lakshminarayana, and Carsten Maple
\thanks{H. Jahangir (Corresponding author), S. Lakshminarayana and  are with the School of Engineering, University of Warwick, UK. C. Maple is with the Warwick
Manufacturing Group, University of Warwick, UK}
\thanks{E-mails: (Hamidreza.Jahangir, Subhash.Lakshminarayana, CM)@warwick.ac.uk }}

\markboth{ }%
{}


\maketitle

\begin{abstract}
Recent research has shown that the security of power grids can be seriously threatened by botnet-type cyber attacks that target a large number of high-wattage smart electrical appliances owned by end-users. Accurate detection and localization of such attacks is of critical importance in limiting the damage. To this end, the paper proposes a novel technique using capsule networks (CNs) tailored to the power grid security application that uses the frequency and phase angle data monitored by phasor measurement units (PMUs). With the benefit of vector output from capsules and dynamic routing agreements between them, CNs can obtain accurate detection and localization performance. To demonstrate the efficiency of the suggested technique, we compare the developed CN with benchmark data-driven methodologies, including two-dimensional convolutional neural networks (2D-CNN), one-dimensional CNN (1D-CNN), deep multi-layer perceptrons (MLP), and support vector machines (SVM). Simulations are performed on IEEE 14-, 39-, and 57-bus systems, considering various real-world issues such as PMU delays, noisy data, and missing data points. The results show that CNs significantly outperform other techniques, thus making them suitable for the aforementioned cyber security applications.
\end{abstract}






\section{Introduction}\label{introduction}
\IEEEPARstart{H}{igh}-wattage bot-net threats compromising of maliciously managed internet-enabled electrical devices (e.g., air conditioners, heaters, electric vehicle charging points, etc.) can pose serious threat to power grid operations. By manipulating a substantial portion of the demand, these attacks can disrupt power grid operations that can result in severe consequences, such as unsafe frequency excursions, line outages and increase the power grid's operational cost \cite{MohsenianLAA, soltan2018blackiot}. Dynamic load altering attacks (D-LAAs), in which an attacker introduces a series of load perturbations over time following the frequency fluctuations, can disrupt power grid control loops \cite{AminiLAA2018, LakshIoT2021}. 
Low-inertia conditions with a high penetration of renewable energy resources can further exacerbate the grid's vulnerability to LAAs \cite{LakshCOVID2022}. 



In order to defend the power grid against such attacks, existing works have focused on both offline security reinforcements and online detection/mitigation. In offline measures, cost-effective load protection strategies (e.g., by enabling tamper-proof/encryption-enabled load devices) that can prevent the destabilizing effects of D-LAAs were investigated in \cite{AminiLAA2018, LakshIoT2021}. However, given the lack of unified security standards in IoT-based electrical appliances, these prevention-based approaches are challenging to implement practically. Thus, online measures to mitigate LAAs (responding to LAAs using generation redispatch) are crucial \cite{CHEN2020116015, GUO2021107113, ZhongdaMitigation2022}. The success of these online measures relies on accurate detection and localization of LAAs.  

To this end, the main approach lies in inferring the attacked nodes using power grid signals monitored by phasor measurement units (PMUs). The authors in \cite{AminiIdentification2019} developed a hierarchical localization task for D-LAAs by analyzing power grid frequency/phase angle data in the frequency domain using an optimization approach. However, the algorithm is developed specifically for a linear system model. Alternatively, the authors in \cite{IzbickiACC2017} framed attack detection/identification as a parameter estimation problem and solved using an unscented Kalman filter (UKF) approach, which however, suffers from scalability issues when considering multi-point attacks. To overcome these limitations, recent work has applied machine learning (ML) techniques \cite{lakshminarayana2021datadriven, Pepin2022}. However, as we show in this work, traditional ML methods such as support vector machines proposed in \cite{Pepin2022} do not provide satisfactory performance in large-scale systems. Other methods such as physics-informed ML technique proposed in \cite{lakshminarayana2021datadriven}  suffer from implementation issues related to combining power grid operational equations with ML's training loss, limiting their applicability (e.g., physics-informed neural network method fares poorly in slow oscillatory systems). Thus, existing literature lacks a robust framework for detecting/localizing LAAs.





ML techniques have also been applied in prior work in the context of fault detection and classification \cite{huang2020synchrophasor, DekaLine2019, noise_li2021power}. However, most of these works apply conventional techniques such as deep multi-layer perceptrons (MLPs)  and convolutional neural networks (CNNs). In addition to the practical limitations of these approaches, {such as performance degradation due to missing/outlier points in the input PMU datasets \cite{PMU}}, they also suffer from significant drawbacks when applied in the context of detecting/localizing attacks as enlisted in the following. (i) In the context of a cyber attack, the exact time corresponding to the onset of an attack cannot be known by the operator. Thus, there is an inevitable delay between the actual instant at which the attack is launched and the data observation time window considered for attack detection/localization. An example is shown in Fig.\ref{fig:delay_effect}. In the ML context, this implies that the online observation profiles used to localize the attack source are not synchronized with the offline training profiles (as shown in Fig.\ref{fig:delay_effect}). The ML classifier must be able to provide accurate localization performance with such data inconsistencies. (ii) In the context of LAAs, multiple combinations of attack parameters result in dynamics that are very similar, as shown and explained in Fig. \ref{fig:similar cases}. In such instances, we require a high-resolution classifier that can differentiate between very similar observation profiles. Our results show that in both these cases, existing MLPs and CNNs do not provide satisfactory performance.

To overcome these drawbacks, we develop a novel data-driven solution tailored to power grid monitoring data to detect/localize LAAs based on capsule networks (CNs). The key feature of CNs that makes them suitable to handle the aforementioned challenges is their ability to capture ``positional instantiations'' (PI), which refers to the position of different features within a dataset. Conventional ML approaches such as CNNs only detect the presence or absence of certain features in the dataset. However, they do not record the exact position of the features within the data. For instance, in the frequency dynamics shown in Fig. \ref{fig:delay_effect}, PI refer to the positioning of each observed point in the profiles in the general space of the input data and with respect to other profiles. Moreover, by implementing vector output of capsules and internal loops in their feedforward information flow (known as dynamic routing agreements), {CNs are capable of capturing and analyzing the position details of the extracted features from observed data. \cite{ribeiro2022learning}.} This provides CNs with the ability to differentiate between system dynamics that are very similar (such as those illustrated in Fig. \ref{fig:similar cases}). We redesign CNs tailored to power grid datasets and ensure their security (\textcolor{black}{more details are provided in Section \ref{subsec: designing network}).}

{Specifically, we examine a short observation window of the frequency/phase angle data monitored by PMUs located at generator bus terminals — just 2 seconds of the examined profiles to be in compliance with the response time of the power grid relays as specified by IEEE Standard 1547 \cite{8332112}. In addition, diverse attacks, including single- and multi-point attacks (see Section \ref{sec:numerical_setting}) that make unstable and semi-unstable states to power grid (see Section \ref{subsec: gathering data}) are considered in this study.}

A summary of the key contributions of this paper is as follows:

\begin{itemize}
\item Introducing a novel online monitoring data-driven technique based on CN architecture that is specifically designed for power grid data – this classifier network can precisely detect and identify the source of LAAs within a short observation window (lower than two seconds). This is, to our knowledge, the first study to include PIs of power grid data to localize the primary source of the impending threats posed by high-wattage IoT-enabled devices.
\item \textcolor{black}{Using vector features (i.e., output layers with 16 and 32 instantiations for IEEE 39- and 57-bus cases, respectively, see Section \ref{capsule_layers}) and considering an internal loop between the capsule layers (i.e., routing agreement algorithm between the Primary and Digit capsules, see Section \ref{Dynamic routing agreements}) contribute to the development of a robust method that can deal with missing and outlier points in PMU profiles, common issues in real-world applications that have significant impacts on the stability of other data-driven methods. }
\item \textcolor{black}{Capturing the PIs of the observed profiles enables CNs to present accurate localization results in challenging circumstances, such as overlapping input data profiles with unexpected delays (training data and test data are not synchronized, as shown in Fig. \ref{fig:delay_effect}). Allocating a unique capsule to each target class in order to verify the presence and arrangement of its PIs in the observed profiles is the primary key factor in delivering highly accurate findings in similar cases (see Section \ref{sec:numerical_simulation_without_noise} and Fig. \ref{fig:similar cases}) and random inevitable delays (see Section \ref{sec:numerical_simulation_delayed} and Fig. \ref{fig:PMU delay}). This feature distinguishes the CN's performance in real-world applications from other techniques.}


\end{itemize}

\section{Problem statement}

\subsection{Power grid dynamics under LAAs}\label{power grid modeling}
To examine the dynamics of the power grid in the presence of LAAs, we consider a standard power grid model whose dynamics can be described by the following differential equations \cite{AminiLAA2018} (\textbf{Nomenclature I} contains complete information on the parameters):
\begin{table}
\centering
\captionsetup{font=footnotesize}
\caption*{\textbf{Nomenclature I}:  Parameters of dynamic  power grid modeling}
\label{tab:nomenclature}
\scalebox{0.75}{%
\begin{tabular}{ll}
\hline
Parameters                                             & Defenitions                                         \\ \hline
$N$, $\mathcal{N} = \mathcal{N}_G \cup \mathcal{N}_L,$ & total bus number                                    \\
$N_G = |\mathcal{N}_G|$                                & set of generator buses                              \\
$N_L = |\mathcal{N}_L|$                                & set of load buses                                   \\
$M$                                                    & total transmission line number                      \\
$\deltav \in \RR^{N_G}$                                & phase angle of the generator buses                  \\
$\thetav \in \RR^{N_L}$                                & phase angle of the load buses                       \\
$\omegav \in \RR^{N_G}$                                & generator bus frequencies                           \\
$\Mm\in  \RR^{N_G \times N_G}$                         & diagonal matrices of generator inertia coefficients \\
$\Dm^G \in  \RR^{N_G \times N_G}$                      & diagonal matrices of damping coefficients           \\
$\Km^I \in  \RR^{N_G \times N_G}$                      & diagonal matrices of proportional coefficients      \\
$\Km^P \in  \RR^{N_G \times N_G}$                      & diagonal matrices of integral coefficients          \\
\multirow{4}{*}{\begin{tabular}[c]{@{}l@{}}$\Bm_{bus} = \begin{bmatrix} \Bm^{GG}  & \Bm^{GL} \\ \\ \Bm^{LG} & \Bm^{LL}\\ \end{bmatrix}$\end{tabular}} &
  \multirow{4}{*}{admittance matrix} \\
                                                       &                                                     \\
                                                       &                                                     \\
                                                       &                                                     \\
$\Bm^{GG} \in \RR^{N_G \times N_G}$                    & sub-matrices of the admittance matrix (Gen-Gen)     \\
$\Bm^{LL} \in \RR^{N_L \times N_L}$                    & sub-matrices of the admittance matrix (Line-Line)   \\
$\Bm^{GL} \in \RR^{N_G \times N_L}$                    & sub-matrices of the admittance matrix(Gen-Line)     \\
$\epsilonv^L \in \RR^L$                                & \emph{static} LAA component                         \\
$\Km^{L} \omegav $                                   & \emph{dynamic} LAA component (D-LAAs)                         \\
$\Km^{L} \in \RR^{N_L} \times \RR^{N_G}$               & matrix of attack controller gain values  \\
$\omegav$                                              & frequency fluctuations of the system                \\
$\pv^L$                                                & total system load                                   \\
$\pv^{LS}$                                             & secured portion of the load                          \\
$\pv^{LV}$                                             & vulnerable portion of the load                      \\ \hline
\end{tabular}%
}
\end{table}

\begin{figure}
    \centering
  \includegraphics[height=7 cm,width=8.5 cm,trim= 15 20 30 58,clip]{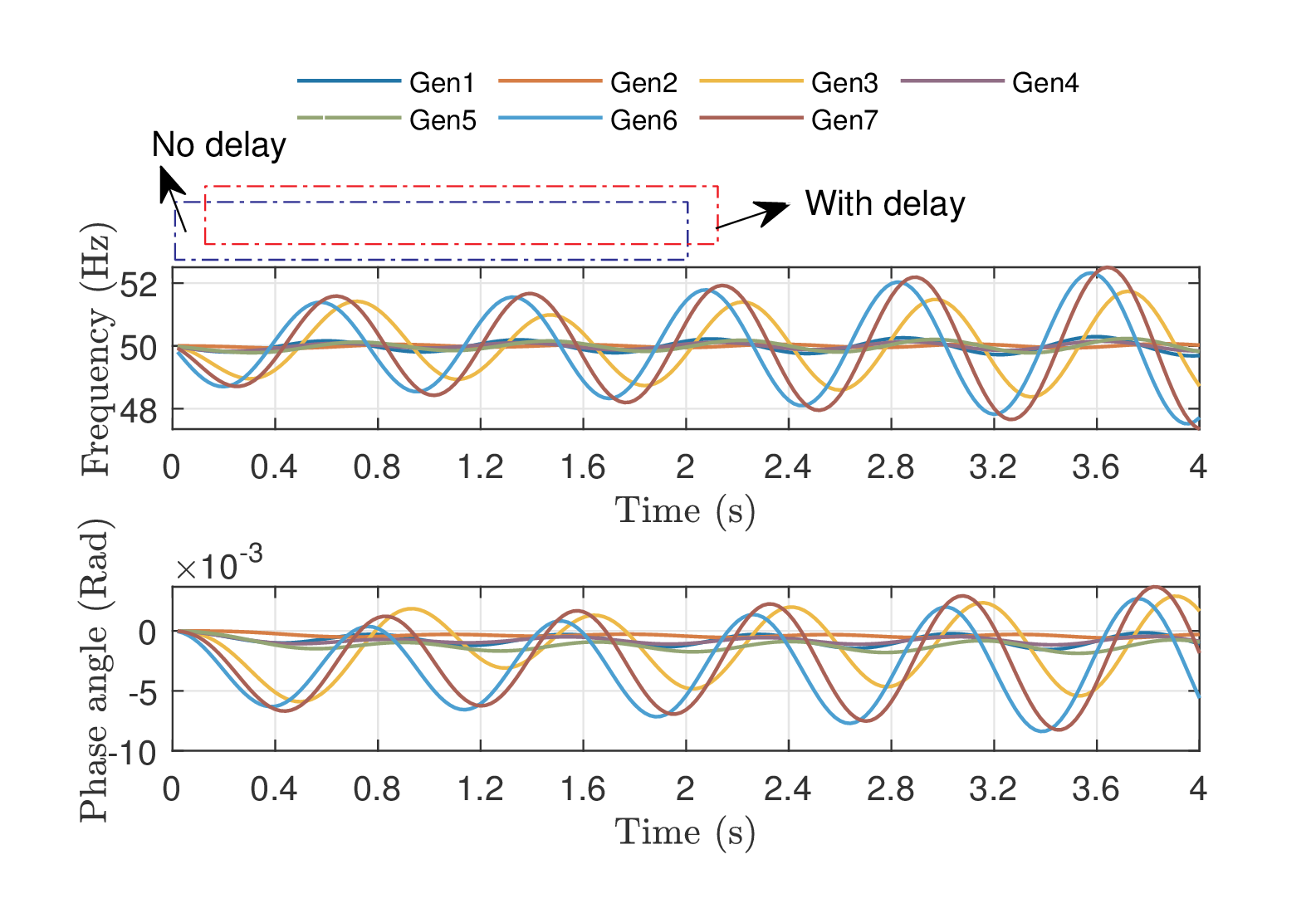}
  \caption{Frequency and phase angle profiles of generator buses in IEEE 57-bus system following the initiation of a
D-LAA with $K_L(39,6)=92.6$ pu and $\epsilon_{24} =0.4$ MW.}\label{fig:delay_effect}
\end{figure}

\begin{align}
& \begin{bmatrix} 
\Id & {\bf 0} & {\bf 0} \\
{\bf 0} & -\Mm & {\bf 0} \\
{\bf 0} & {\bf 0} & {\bf 0}
\end{bmatrix}
\begin{bmatrix} 
\dot{\deltav} \\
\dot{\omegav} \\
\dot{\thetav}
\end{bmatrix} = \begin{bmatrix} 
{\bf 0} \\
{\bf 0} \\
\pv^{LS} + \epsilonv^L
\end{bmatrix} + \nonumber \\
& \begin{bmatrix} 
{\bf 0} & \Id & {\bf 0} \\
\Km^I + \Bm^{GG} & \Km^P + \Dm^G  & \Bm^{GL} \\
\Bm^{LG} & -\Km^L & \Bm^{LL}
\end{bmatrix}
\begin{bmatrix} 
{\deltav} \\
{\omegav} \\
{\thetav} 
\end{bmatrix}. 
\label{eqn:dyn_mtx}
\end{align}

\noindent To characterize the LAAs, we first split the power grid load ($\pv^L$) of each bus into two clusters as follows:
\begin{align}
   \pv^L = \pv^{LS} + \pv^{LV}.
   \label{eqn:main}
\end{align}

\noindent The safe portion of the load ($\pv^{LS}$) excludes smart high-wattage devices or is well guarded, while the vulnerable portion of the load ($\pv^{LV}$) is comprised of IoT-enabled high-wattage devices such as smart electric vehicle charging stations, thermostats, etc. We delve deeper into the $\pv^{LV}$ to model LAAs as follows:
\begin{align}
  \pv^{LV} =  \epsilonv^L - \Km^{L} \omegav.
  \label{eqn:devided}
\end{align}
\noindent $\epsilonv^L$ refers to \textit{static} LAA in which the attacker imposes a step-change to the power grid's load (i.e., turning the manipulated devices on or off abruptly and without further alteration). $-\Km^{L} \omegav$, on the other hand, relates to the \textit{dynamic} LAA, in which the attacker observes the frequency of the sensing buses ($\omegav$) and attempts to behave adversely in response to frequency fluctuations (i.e., the attacker alters the load in the reverse way of the frequency deviations). $\Km^{L} \in \RR^{N_L} \times \RR^{N_G}$ is a matrix of attack controller gain values that specifies which gain value is assigned to each sensing bus. Finally, the ultimate values of LAAs are restricted as follows:
\begin{align}
 {\Km}^{L}\omega   \leq (P^{LV} - \epsilon^L)/2 \label{eqn:Attack_lim}.
\end{align}
Fig.\ref{fig:delay_effect} depicts a sample of the aforementioned attacks on the IEEE 57-bus system in order to provide a better understanding of the fluctuations in power grid frequency and phase angle owing to the imposed LAA (attack vectors are given in the caption part of Fig.\ref{fig:delay_effect}).

\subsection{Monitoring the power grid for LAAs localization}\label{monitoring}
With PMUs distributed throughout the power grid, the Independent System Operator (ISO) continually monitors the power grid's frequency $\{ \dot{\delta}^{(\tau)}_i \}_{i \in \mathcal{N}_G, \tau = 1,\dots,T}$ and phase angle $\{ \delta^{(\tau)}_i \}_{i \in \mathcal{N}_G, \tau = 1,\dots,T}$ of the various points. Fig.\ref{fig:delay_effect} illustrates an example of the observed data from the installed PMUs in generator buses. $x^{(\tau)}$ represents the value of the signal $x$ at time slot $\tau$, with slots sampled at a time interval of $T_s$ and $T$ representing the total number of time slots. IEEE/IEC standards stipulate that the PMU sampling frequency for a 50 Hz system may range between $10$ and $100$ frames per second.
Thus, $T_s$ is between $10-100$ ms \cite{PMU2018}.

In this study, we identify and pinpoint the bus from which cyberattacks are launched. We monitor the frequency $\{ {\omega}^{(\tau)}_i \}_{i \in \mathcal{N}_G, \tau = 1,\dots,T}$ and phase angle $\{ \delta^{(\tau)}_i \}_{i \in \mathcal{N}_G, \tau = 1,\dots,T}$ profiles of PMUs for this purpose. Specifically, we want to identify the locations associated with non-zero values of $\epsilonv$ and $\Km^L$ elements for destabilizing attacks (see Section \ref{subsec: designing network} for more details).

\section{Designing capsule networks for monitoring power grid} \label{desiging capsules}

\subsection{Brief Introduction to Capsule Nets}\label{why capsule nets}



\begin{figure}
    \centering
  \includegraphics[height=7 cm,width=9 cm,trim= 300 110 220 50,clip]{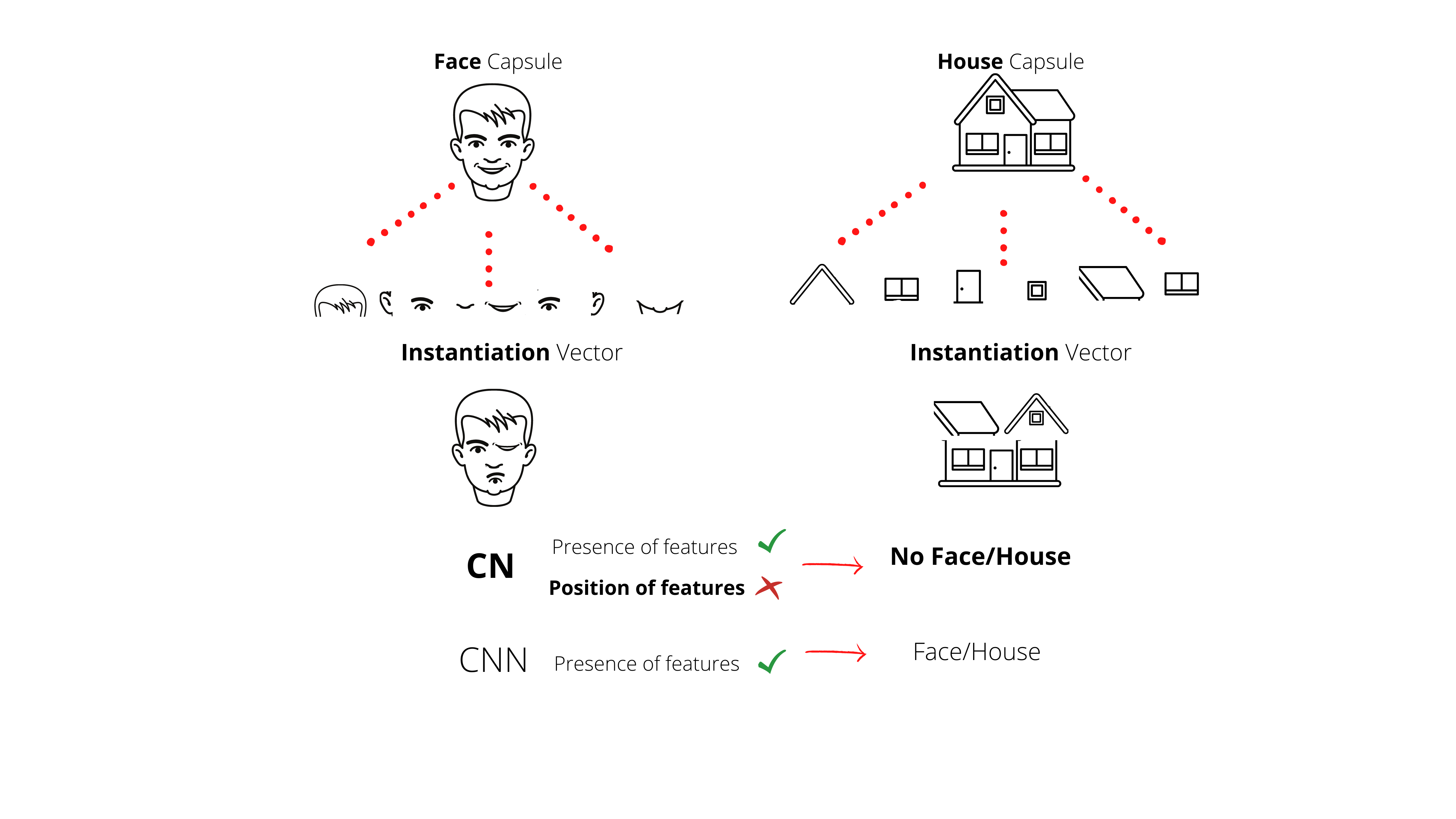}
  \caption{General concept of the capsules and their instantiation.}\label{fig:General idea}
\end{figure}

Despite being robust classifiers, CNNs suffer from some drawbacks in classification tasks, as outlined by Hinton \cite{sabour2017dynamic,ribeiro2022learning}. An illustration is provided in Fig.\ref{fig:General idea}, where we present two versions of a human face (and a house), which have identical features, but the features are placed at different positions/orientations within the respective images. When mentioned samples are presented as inputs to a CNN, its hidden layers will identify the feature sets in the images (presented at the center). However, the CNN will not be able to distinguish between the top and bottom versions of the face (and the house) since both the figures have identical features. This is due to the following reasons. (i) The hidden layers of CNNs detect the presence or absence of features with an image (or a given dataset). They do not record the position or orientation of a feature set within the dataset. This is because of the pooling layers (e.g., max-pooling and/or average-pooling) that eliminate the positional features (instantiations) of the observed profiles as they only capture the maximum and mean value of the monitored regions within the image \cite{CapsuleCNN}. (ii) To obtain a generalized representation, CNNs should be trained on a large dataset containing images with all possible orientations. However, this demands significant computational power and memory requirements. {This is the \emph{invariance problem}; CNNs cannot deal with rotational invariance (i.e., examples from different points of view that are entirely new samples for them and are not generated by augmentation methods) \cite{xi2017capsule}.}

In contrast, CNs implement the so-called \emph{capsules}, which are a collection of vector neurons rather than scalar neurons used in the case with standard neural networks, such as CNNs \cite{Capsule_Cyber}. \textcolor{black}{As depicted in Fig. \ref{fig:general capsule}, each capsule has $N (\geq 1) $ dimensions that correspond to its internal vector-based instantiations (i.e., position, size, and orientation of the captured elements, which can be defined as the nose, ears and eyes shown in Fig. \ref{fig:General idea}), whereas a typical neuron has only one scalar component.} In CNs, one particular capsule from the last capsule layer (called the ``Digit capsule'' layer) is assigned to each target class in order to check the presence of the mentioned instantiations of that target class. In addition, CNs are equipped with an internal computation called \emph{dynamic routing} agreement that determines which capsule should be triggered based on the captured attributes from input data (described in Section \ref{Dynamic routing agreements}). These characteristics enable us to avoid any misclassification, as depicted in Fig. \ref{fig:General idea}.

\subsection{{Application of Capsule Nets to LAA Detection/Localization}}
In the context of detection/localization of D-LAAs, we leverage the voltage phase angle and frequency data at the generator buses that are monitored by the PMUs. As depicted in Fig.\ref{fig:delay_effect}, the instantiation parameters of the datasets include the amplitude and position of the observed fluctuations. The application of CNs for the detection/localization tasks examines both the presence and position of the features within the total time window, as well as their position relative to each other. 


As noted before, in cyber security applications, a significant challenge is that the exact time corresponding to the onset of the attack cannot be known by the system operator. This can present challenges in the application of ML, as it can result in an offset between the datasets corresponding to the training and testing phases. 
An example is illustrated in Fig.~\ref{fig:delay_effect}, where we present data sets from two different observation windows. In particular, to simulate the effect of the offset, we train a neural network (during the offline phase) with a dataset corresponding to power grid dynamics generated following the injection of a cyber attack (i.e., the dataset from $0-2$ seconds interval). However, during the online operations, we test the classifier's performance with a dataset from the interval $0.5 - 2.5$ seconds, wherein the offset is used to represent the detection delay. In such cases, we require a robust deep learning network capable of handling the classification task from different orientations (i.e., in this case, offset between the train and the test data). This unique characteristic is possible by CNs due to their investigation of data in vector space, which enables them to provide superior performance in these circumstances, as proved in image processing tasks \cite{sabour2017dynamic}.
Dynamic routing between capsules permits this network to send the relevant attributes of the captured entity to the appropriate parents of the preceding layer. To summarize, CNs replace CNNs' scalar output descriptors and max-pooling layers with vector output capsules and dynamic routing agreements, respectively. In contrast to the max-pooling layers, which discard the position information of the captured entity, these details are retained in capsules and leveraged in the feature extraction process by dynamic routing agreements \cite{CapsuleCyber}. This is extremely helpful for the highly overlapping input data; as shown in Fig.\ref{fig:delay_effect}, our input data, in which we observe considerable overlap of the frequency and phase angle profiles.

\subsection{Gathering data for our supervised classifier} \label{subsec: gathering data}
In this study, we consider the captured frequency/phase angle profiles (2 features) from PMU devices at generator buses ($N_G$) as input data, shown in Fig.\ref{fig:delay_effect}. Consequently, the arrangement of our input data will be $N_G\times T \times 2$ -- T represents the number of time steps associated with each observation sample. In this work, the sampling interval is 20 milliseconds, yielding 50 samples every second observation. Diverse D-LAA types are investigated in order to enrich our training data for our supervised classifier using Monte Carlo simulation techniques on the understudied cases (i.e., IEEE 14-, 39-, and 57-bus systems). Next, we explore the eigenvalues (defining them by the $a+bj$ structure) of the power grid (in the state-space representation) to determine if the implemented load-altering efforts are cyberattacks (with at least one eigenvalue in the right-half plane, i.e. $a>0$, which transforms the power grid into an unstable state) or ordinary power grid oscillations (without any eigenvalue in the right-half plane)\cite{kundur2007power}. Moreover, according to the North American Electric Reliability Corporation, when $\zeta\leq 3\%$, $2.5\leq \omega _n\leq 12.6$ rad/s (the eigenvalues of the power grid can be stated in terms of damping ratio $\zeta$ and natural oscillation frequency $\omega _n$: $a = -\zeta\omega _n , b = \omega _n \sqrt[]{1-\zeta^2}$  \cite{tabrizi2017power}), the power grid will be in a semi-unstable state; in this situation, the power grid will be unstable due to its inherent oscillations. Our training library also takes into account these types of attacks.

\section{Overall Structure of the Designed Capsule Net} \label{subsec: designing network}


\begin{algorithm}[tbh]
\scriptsize
	\caption{Capsule Net for \textbf{IEEE 39-and 57-bus cases}}
    \underline{Input:} Phase angle data $\{ \delta^{(\tau)}_i \}_{i \in \mathcal{N}_G, \tau = 1,\dots,T}$ and frequency data $\{ \omega^{(\tau)}_i \}_{i \in \mathcal{N}_G, \tau = 1,\dots,T}$, where, for IEEE 39-bus case, $\mathcal{N}_G = \{30,\dots,39\},  |\mathcal{N}_G| = 10$ and for IEEE 57-bus case, $\mathcal{N}_G = \{1,2,3,6,8,9,12\}, |\mathcal{N}_G| = 7.$\\
    \underline{Labels:} Defining the location of the attacked bus as labels  ( $i \in \mathcal{N}_L$), where, for IEEE 39-bus case $ |\mathcal{N}_L| = 29.$ and for IEEE 57-bus case $|\mathcal{N}_L| = 50.$ \\
    \underline{Defined parameters:} $T=100, W_{Conv1}, b_{Conv1}, W_{Conv1}, b_{Conv2}, W_{p,q}$ = randomly made based on Gaussian distribution\\
    
	\begin{algorithmic}[1]
  \FOR{ $n_{batch}$ iteration}
      \STATE Define $S$ as input data vector,\\
      $S\in \Re ^{10\times 100\times 2}$ (\textbf{IEEE 39-bus})\\
      $S\in \Re ^{{7}\times 100\times 2}$. (\textbf{IEEE 57-bus})
      \STATE  Apply the first 2D-CNN layer by \text{512} Kernels with size ($\text{1}\times\text{10}$)  and stride ($\text{1}\times\text{10}$):\\\textit{}
      $Z_{Conv1} \gets ReLU(W_{Conv1}\otimes S+b_{Conv1})$\\ 
      ${Z}_{Conv1}\in \Re ^{10\times 10\times512}$ (\textbf{IEEE 39-bus})\\
      $Z_{Conv1}\in \Re ^{7\times 10\times 512}$(\textbf{IEEE 57-bus}) \\
      
      \STATE Pass dropout layer (\text{0.1})\\
      \STATE  Apply the second 2D-CNN layer by \text{256} Kernels with size ($2\times2$) / (${2}\times{5}$)  and stride ($2\times2$) / (${1}\times{2}$): (\textbf{IEEE 39-bus})/(\textbf{IEEE 57-bus})\\
      $Z_{Conv2} \gets ReLU(W_{Conv2}\otimes Z_{Conv1}+b_{Conv2})$\\ 
      $Z_{Conv2}\in \Re ^{5\times 5\times256}$ (\textbf{IEEE 39-bus})\\
      $ Z_{Conv2}\in \Re ^{ {5}\times {5}\times{256} }$ (\textbf{IEEE 57-bus})\\
      \STATE Reshape: $Z^{P_{Caps}} \in \Re^{800\times 8 \times 1} \gets Z_{Conv2}\in \Re ^{5\times 5\times 256}$ (\textbf{IEEE 39-bus})\\
      \text{Reshape}: $Z^{P_{Caps}}\in \Re^{576\times 16\times 1} \gets {Conv2}\in \Re ^{\text{6}\times 6\times256}$  (\textbf{IEEE 57-bus})\\
      \STATE Pass data from Primary Caps to internal vectors:\\ $Z_{q\mid p} \gets W_{pq}\, Z_{p}^{P_{Caps}}$\\
      (\textbf{IEEE 39-bus})\\
      $Z_{p}^{P_{Caps}}\in \Re ^{8\times 1} \subset Z^{P_{Caps}}\in \Re ^{800\times 8\times1}$, $_{p = 1,\dots,800}$\\
      $W_{pq}\in \Re ^{8 \times 16}$ \\
      $Z_{q\mid p} \in \Re^{1 \times 16} \subset Z^{Digit_{Caps}} \in \Re^{29 \times 16}$, $_{q =1,\dots,29}$\\
      (\textbf{IEEE 57-bus})\\
      $Z_{p}^{P_{Caps}}\in \Re ^{\text{16}\times \text{1}} \subset Z^{P_{Caps}}\in \Re ^{\text{576}\times \text{16}\times\text{1}}$, $_{p = \text{1},\dots,\text{576}}$\\
      $W_{pq}\in \Re ^{\text{16}\times \text{32}} $ \\
      $Z_{q\mid p} \in \Re^{\text{1} \times \text{32}} \subset Z^{Digit_{Caps}} \in \Re^{\text{50} \times \text{32}}$, $_{q = \text{1},\dots,\text{50}}$\\
      \STATE Define zero values for $b_{pq}$
      \PROCEDURE{Dynamic routing}{$r,b_{pq},Z_{q\mid p}$}
      \FOR{$r$ iteration}
      \STATE Calculate coupling parameters:\\
      $c_{pq} \gets \exp (b_{pq}) /\sum_{k=1}^{K}\exp(b_{pk}))$\\ 
      $_{k = 1,\dots,29}$ (\textbf{IEEE 39-bus}), $_{k = \text{1},\dots,\text{50}}$ (\textbf{IEEE 57-bus}) \\
      $b_{pq},c_{pq} \in \Re ^ {800 \times 29}$ (\textbf{IEEE 39-bus})\\
      $b_{pq},c_{pq} \in \Re ^ {\text{576} \times \text{50}} (\textbf{IEEE 57-bus})$
      \STATE Calculated weighted sum of internal vectors:\\
      $d_{q}  \gets \sum_{p}^{}c_{pq}Z_{q\mid p}$\\
      $d_{q} \in \Re ^ {16\times 1}$ (\textbf{IEEE 39-bus})\\
      $d_{q} \in \Re ^ {\text{32} \times \text{1}} $ (\textbf{IEEE 57-bus})
      \STATE Calculate output of Digit Caps:\\
      $v_{q}^{Digit_{Caps}} \gets Squashing (d_{q})$\\
      $v_{q} \in \Re ^ {16 \times 1}$ (\textbf{IEEE 39-bus})\\
      $v_{q} \in \Re ^ {\text{32} \times \text{1}} $ (\textbf{IEEE 57-bus})
      \STATE Updating $b_{pq}$ (agreement process):\\
      $b_{pq} \gets b_{pq} + Z_{q\mid p} v_{q}$
      \ENDFOR
      \ENDPROCEDURE
      \STATE\textbf{return} $v_{q}$
      \STATE Calculate marginal loss function ($L_{M}$) as sum of $L_q$ (based on Equation \ref{eqn:marginal loss})
      \STATE Updating trainable parameters\\
      $ W_{Conv1} \gets W_{Conv1} + \partial L_{M}/\partial W_{Conv1}$\\
      $ b_{Conv1} \gets b_{Conv1} +  \partial L_{M}/\partial b_{Conv1}$\\
      $ W_{Conv2} \gets W_{Conv2} + \partial L_{M}/\partial W_{Conv2}$\\
      $ b_{Conv2} \gets b_{Conv2} + \partial L_{M}/\partial b_{Conv2}$\\
      $ W_{pq} \gets W_{pq} + \partial L_{M}/\partial W_{pq}$
  \ENDFOR
\end{algorithmic}
\label{algorithm}
\end{algorithm}

\begin{table}
\centering
\captionsetup{font=footnotesize}
\caption*{\textbf{Nomenclature II}:  Parameters of CN  (dimension are given in \textbf{Algorithm \ref{algorithm}})}
\label{tab:nomenclature2}
\scalebox{0.75}{%
\begin{tabular}{ll}
\hline
Parameters                                             & Definitions                                         \\ \hline
$S$                                                    & input data vector                                   \\
$ b_{Conv1}, b_{Conv2}$                                & bias vector of first and second convolutional layer  \\
$ W_{Conv1}, W_{Conv2}$                                & weight matrix of first and second convolutional layer \\
$ Z_{Conv1}, Z_{Conv2}$                                & output vector of first and second convolutional layer \\
$Z^{P_{Caps}}$                                         & vector of Primary capsules                           \\
$Z^{Digit_{Caps}}$                                     & vector of Digit capsules                              \\
$p,q$                                                  & index of Primary and Digit capsules                  \\
$Z_{p}^{P_{Caps}}$                                     & vector of p\textit{th} Primary capsules                              \\
$Z_{q}^{Digit_{Caps}}$                                 & vector of p\textit{th} Digit capsules                              \\
$W_{pq}$                                               & transition matrix between the Primary and Digit capsules \\
$Z_{q\mid p}$                                          & internal vector between p\textit{th} Primary and q\textit{th} Digit capsules  \\
$r$                                                    & number of iterations for dynamic routing algorithm\\
$b_{pq}$                                               & beginning logits between p\textit{th} Primary and q\textit{th} Digit capsules\\
$c_{pq}$                                               & coupling coefficients between p\textit{th} Primary and q\textit{th} Digit capsules \\
$d_{q}$                                                & output of q\textit{th} Digit capsule before applying squash function\\
$v_{q}$                                                & output of q\textit{th} Digit capsule after applying squash function \\
$L_q$                                                  & marginal loss value of q\textit{th} Digit capsule      \\
$L_M$                                                  & total marginal loss of all Digit capsules      \\
 \hline
\end{tabular}%
}
\end{table}
The proposed CN structure is shown in Fig.~\ref{fig:Overall structure of capsule Nets}, and the related notations are specified in the Nomenclature Table for CN modeling.
The network structure is tailored to the context of detection and localization of the D-LAAs; it differs from the CN architecture originally proposed in \cite{sabour2017dynamic}, which was primarily designed for the image classification task. In particular, we have two distinct power grid physical signals (i.e., frequency and voltage phase angle profiles) that have values in different scales, as opposed to image datasets where all the streams have similar scales/characteristics. Thus, using an encoder at the end of the capsule layers (as introduced in \cite{sabour2017dynamic}) applied to the power grid dataset under consideration causes convergence issues, including vanishing/exploding gradients. To stabilize the training procedure, we exclude the encoder in our CN design\footnote{Note that the encoder was introduced in \cite{sabour2017dynamic} only to reconstruct the input image in order to demonstrate that all the features required to reconstruct the input data are stored in the capsules}. Furthermore, to reshape the dataset (which is rectangular and consists of two different data streams of varying scales), we include additional convolutional and dropout layers before the capsule layers (to be explained in the following). Other parameters, such as the layout of the convolutional layers (kernel size, stride values) and capsule layers (components of Primary and Digit capsules), have been meticulously designed for our case studies, presented in \textbf{Algorithm \ref{algorithm}} (\textbf{Nomenclature II} contains complete information on the parameters). The description of the proposed framework is divided into the following four categories (detailed explanations are provided in accordance with \textbf{Algorithm \ref{algorithm}} and Fig. \ref{fig:general capsule}):

\begin{figure}
    \centering
  \includegraphics[height=3.5 cm,width=8.5 cm,trim= 10 180 10 50,clip]{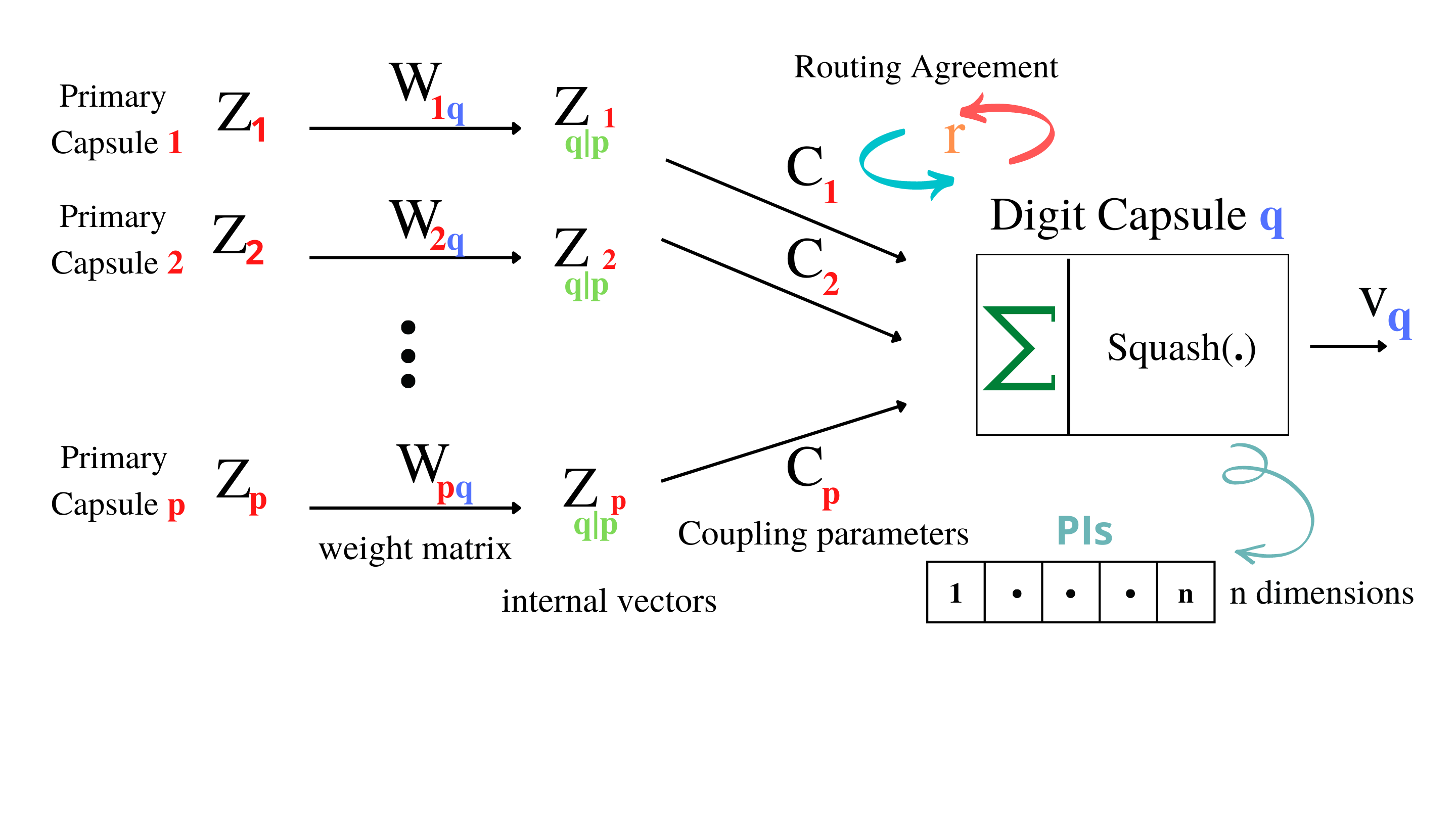}
  \caption{Data flow diagram for a Digit capsule.}\label{fig:general capsule}
\end{figure}





\subsubsection{Initial interpretation of input data using CNN Layers}\label{Data interpretation} 
\begin{figure*}
    \centering
  \includegraphics[height=4.8 cm,width=16.2 cm,trim= 40 290 150 30,clip]{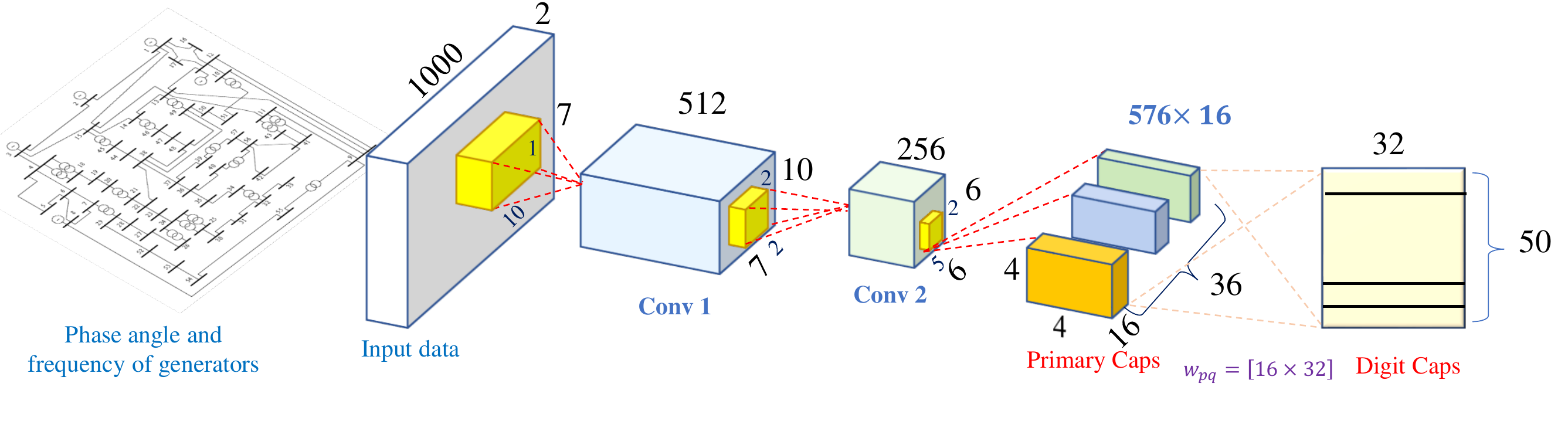}
  \caption{Overall structure of designed CN for IEEE 57-bus case.}\label{fig:Overall structure of capsule Nets}
\end{figure*}

\textcolor{black}{We first preprocess the frequency/phase angle data into a structure that is suitable as inputs to the capsule layers (Primary and Digit capsules are the terms for the initial and output capsule layers, respectively, see Fig.\ref{fig:Overall structure of capsule Nets}). To this end, we add cascaded convolutional layers with ReLU activation functions and without the pooling layers (lines 2–5 of \textbf{Algorithm \ref{algorithm}})}. 
Thus, the input data, which originally consists of a dataset of dimensions $\Re ^{\textbf{10}\times \textbf{100}\times\textbf{2}}$ and $\Re ^{\textbf{7}\times \textbf{100}\times\textbf{2}}$ (triangular-structured, \textcolor{black}{recall its configuration as outlined in Section \ref{subsec: gathering data}}) is transformed to $\Re ^{\textbf{5}\times \textbf{5}\times\textbf{256}}$ and $\Re ^{\textbf{6}\times \textbf{6}\times\textbf{256}}$ (square structure) for the IEEE 39- and 57-bus systems, respectively. Note that in the aforementioned convolutional structure, max-pooling layers are excluded as they disregard the location information of the features.



\subsubsection{Capsule Layers} \label{capsule_layers}

The output of the last convolutional layer ($Z_{Conv2}$) is a stack of featured maps, which serve as inputs to the capsule layers (after reshaping). CNs consist of two types of capsule layers, called ``Primary'' and ``Digit'' capsules, as depicted in Fig. \ref{fig:Overall structure of capsule Nets}. In essence, each Primary capsule recognizes a segment of the objects and, via an internal loop, activates the most relevant Digit capsules in the next layer (also called dynamic routing, explained in the following).

As given in line 6 of \textbf{Algorithm \ref{algorithm}}, for the Primary capsule layer, in the IEEE $39-$bus system, and IEEE $57-$bus system, we have $Z^{P_{Caps}}\in \Re ^{\textbf{800}\times \textbf{8}\times\textbf{1}}$ (800 capsules, where each capsule is a vector of 8 dimensions), and $Z^{P_{Caps}}\in \Re ^{\textbf{576}\times \textbf{16}\times\textbf{1}}$ (576 capsules, where each capsule is a vector of 16 dimensions),  respectively. Note that these dimensions are chosen to provide the best performance on the respective datasets. Then, as shown in Fig. \ref{fig:general capsule}, the data from the Primary capsules ($Z^{P_{Caps}}$) are propagated to the internal vectors ($Z_{q\mid p}$) with the transition matrix ($W_{pq}$) before it reaches the Digit capsules ($v_{q}^{Digit_{Caps}}$) by coupling coefficients ($c_{pq}$)  (line 7 of \textbf{Algorithm \ref{algorithm}}). Each Digit capsule is assigned to a specific target class; for instance, in the IEEE 39-bus case, there are 29 load buses that can present as the source of the D-LAAs, and 29 Digit capsules are allocated to this case. For more complex cases, capsules with greater dimensions are needed \cite{sezer2019capsule}, therefore, $v_{q}^{Digit_{Caps}}\in \Re ^{\textbf{16}\times\textbf{1}}$ and $v_{q}^{Digit_{Caps}}\in \Re ^{\textbf{32}\times\textbf{1}}$ are defined for IEEE 39- and 57-bus, respectively.




\subsubsection{Dynamic routing agreements} \label{Dynamic routing agreements} As shown in  Fig. \ref{fig:general capsule}, the weight matrix ($W_{pq}$) connects capsules in the Primary and Digit layers. However, the agreements between the capsules determine the final classification results. In other words, by determining whether to send more or fewer features to each capsule of the next layer, each Primary capsule attempts to identify the output of the Digit capsules. To construct the coupling coefficients ($c_{pq}$), beginning logits ($b_{pq}$) are specified as zero initially (line 8 of \textbf{Algorithm \ref{algorithm}}), and then the Softmax function is applied to them (line 11 of \textbf{Algorithm \ref{algorithm}}). It should be noted that the coupling coefficients between each Primary capsule and all of the Digit capsules accumulate to one owing to the Softmax function. The weighted sum of internal vectors and coupling parameters is then computed ($d_{q}$) (line 12 of \textbf{Algorithm \ref{algorithm}}). As a result, the coupling parameters, which are the fundamental apparatus of dynamic routing agreements between capsules, will impact the output of the capsules in the Digit layer. The Squash function is then applied to the internal vectors to determine the final output of Digit layer capsules (line 13 of \textbf{Algorithm \ref{algorithm}}):
\begin{align}
   v_q = \frac{\left \| d_q \right \|^{2}}{1+\left \| d_q \right \|^{2}}\frac{d_q}{\left \| d_q \right \|}.
   \label{eqn:squash_function}
\end{align}
We utilize the squash function to verify that no vector's length is more than one since the length of the vector denotes the probability that the given vector is identified correctly \cite{hinton2018matrix}. The agreement procedure occurs at line 14 of \textbf{Algorithm \ref{algorithm}}, where the dot product of internal vectors and the output of Digit capsules are taken to update the initial logits. 

Since the Primary capsules only output to the proper capsule in the Digit layer, these capsules get cleaner input signals and can pinpoint the location of an attack launched against the power grid more precisely. This capability is especially useful in situations with overlapping features, such as in our application, where the frequency and phase angles of the generators' profiles overlap, shown in Fig. \ref{fig:similar cases}. Notably, transformation matrices ($W_{pq}$) are not updated during the dynamic routing agreement because doing so would mislead the gradient descent method  \cite{hinton2018matrix}; dynamic routing agreement occurs during the feed-forward procedure, and trainable parameters such as transformation matrices ($W_{pq}$) are only updated during the back-propagation task using gradient descent. Dynamic routing agreement is an internal voting procedure (i.e., repeats in a feed-forward manner for $r$ times as shown in line 10 of \textbf{Algorithm \ref{algorithm}}) amongst capsules to govern the delivery of additional or fewer features to each capsule based on the classification target's agreement or disagreement. In our application, we set $r =5$.

\subsubsection{Training procedure} \label{training procedure} All trainable parameters (i.e., $W_{Conv1}$, $b_{Conv1}$, $W_{Conv2}$, $b_{Conv2}$, and $W_{pq}$) are adjusted according to the gradient descent approach throughout the training process (line 18-19 of \textbf{Algorithm \ref{algorithm}}). The marginal loss function is implemented as follows for each capsule in the Digit layer \cite{sabour2017dynamic}:
\begin{align}
   L_q = T_q \, max(o,m^+ - \left \| v_q \right \|)^2+ \nonumber \\\lambda \, (1-T_q) max (0,\left \| v_k \right \| - m^-)^2,
   \label{eqn:marginal loss}
\end{align}
where, $m^+ = 0.9$, $m^- = 0.1$, $\lambda = 0.5$, and if class $q$ is presented, $T_q = 1$, otherwise  $T_q = 0$.
The total error is then computed by adding the errors of each capsule.

\section{Simulation Results}\label{sec:numerical}
\subsection{Simulation Settings}\label{sec:numerical_setting}

\begin{figure}
    \centering
  \includegraphics[height=4 cm,width=9 cm,trim= 5 0 5 5,clip]{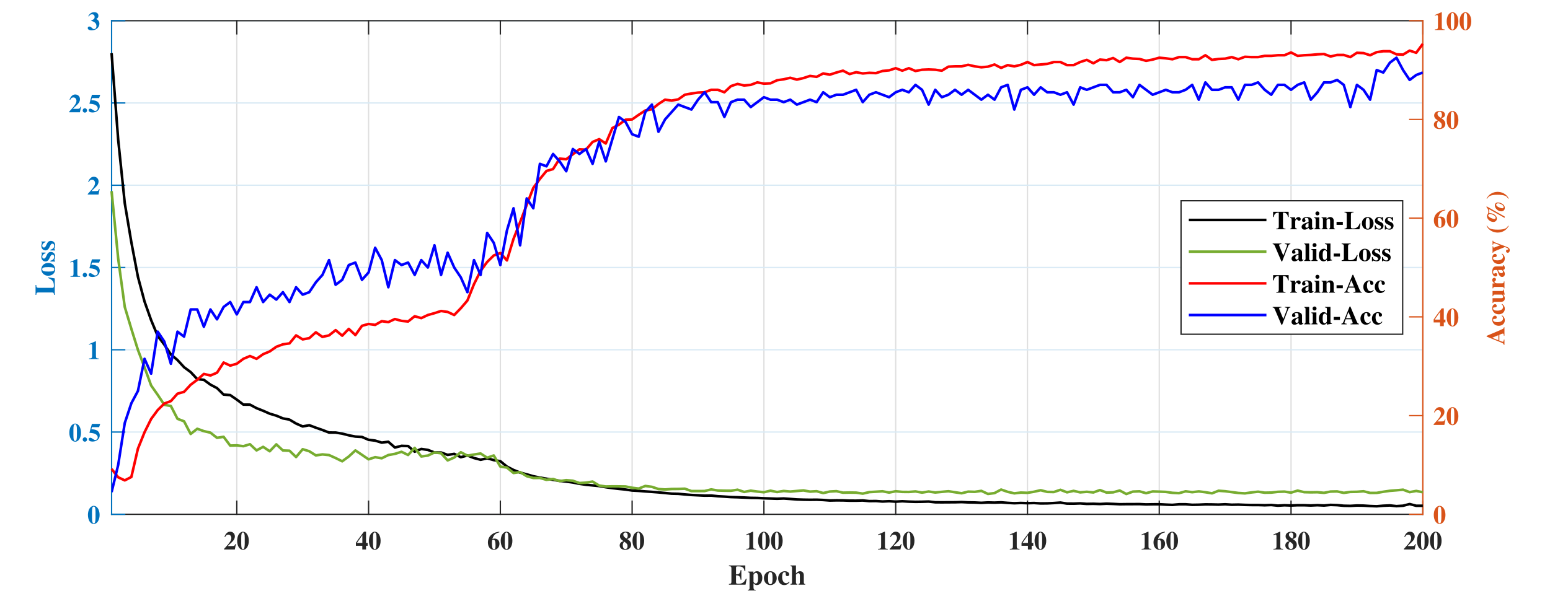}
  \caption{Training plot of CN for IEEE 57-bus system}\label{fig:training plot}
\end{figure}

Three IEEE cases, including IEEE 14-, 39-, and 57-bus systems, are explored to validate the efficiency of the proposed method. \textbf{Algorithm \ref{algorithm}} provides specifics regarding our developed CN for the IEEE 39- and 57-bus systems. However, for the IEEE 14-bus system, we have 9 load and 5 generator buses (i.e., we have $\Re ^{\textbf{5}\times \textbf{100}\times\textbf{2}}$ input data), and for the Primary and Digit layers, we assume 8- and 16-dimensional capsules, respectively. Tracking \textbf{Algorithm \ref{algorithm}} can reveal further information about the intended CN framework for the IEEE 14-bus system. Topological data for these cases are taken from the MATPOWER simulator, and dynamic simulation settings are also included in the Appendix section. In this approach, diverse situations, including delayed, noisy, and missed PMU data samples, are investigated in single-point — the same spot for static and D-LAAs — and multi-point — different locations for static and D-LAAs, and the source of the D-LAA is explored — modes. As stated in Section \ref{introduction}, the detection of D-LAAs has been explored in a variety of research literature; thus, in this study, we present comprehensive results on the localization of D-LAAs. On average, the proposed CN technique achieves greater than 98\% detection accuracy in various settings. Python 3.9 with the TensorFlow framework on a Windows PC with an 11th Generation Intel(R) Core (TM) i7-1185G7 @ 3.00GHz CPU and 16GB RAM was used to develop and test the proposed method.
\textcolor{black}{To train, validate, and test the proposed method, 2000 samples are generated (based on the specified structure in Section \ref{subsec: gathering data}), and the data is segmented as follows: 80\% for training, 10\% for validation, and 10\% for testing. Fig. \ref{fig:training plot} depicts training profiles, including accuracy and total loss, for the validation and training data sets, confirming the stability of our designed CN.} Due to the internal loop of the dynamic routing agreements, the offline training procedure for the proposed method is longer than conventional 2D-CNN methods; however, the online inference of the proposed method, which is a crucial factor in the real-world application of the proposed method, is less than 100 milliseconds, which is an acceptable response rate for detecting and localizing upcoming attacks. \textcolor{black}{Table \ref{tab:time} provides information on the online response time of CNs and other benchmark techniques.}

To demonstrate the promising performance of the CN in localization of the D-LAAs, the numerical findings include a thorough comparison with benchmark machine learning techniques, including 2D-CNN, 1D-CNN, MLP, and SVM (references \cite{zhang2020deep} and \cite{han2021short} provide detailed formulations of CNN networks and SVM classifiers, respectively). Finally, the localization task's accuracy is computed as follows:\cite{dong2021deep}:
\begin{table}
\centering
\caption{Online response time of different approaches}
\label{tab:time}
\scalebox{0.8}{%
\begin{tabular}{cc}
\hline
Method & Online response time (msec) \\ \hline
CN     & 96                          \\
2D-CNN & 79                          \\
1D-CNN & 69                          \\
MLP    & 66                          \\
SVM    & 61                          \\ \hline
\end{tabular}%
}
\end{table}

\begin{align}
  accuracy = \frac{1}{T_n}\sum_{i=1}^{T_n}\frac{C^i}{Q^i},
   \label{eq:accuracy}
\end{align}
where, $C^i$ is the number of correctly classified samples and $Q^i$ represents the number of query examples in the $i_{th}$ test match, and $T_n$ represents the total number of test episodes.

\subsection{Simulation Outcomes and Analysis}\label{sec:numerical_simulation}
\subsubsection{Single-point and multi-point D-LAAs}\label{sec:numerical_simulation_without_noise}
\begin{figure*}
    \centering
  \includegraphics[height=7.5 cm,width=15 cm,trim= 20 10 30 140,clip]{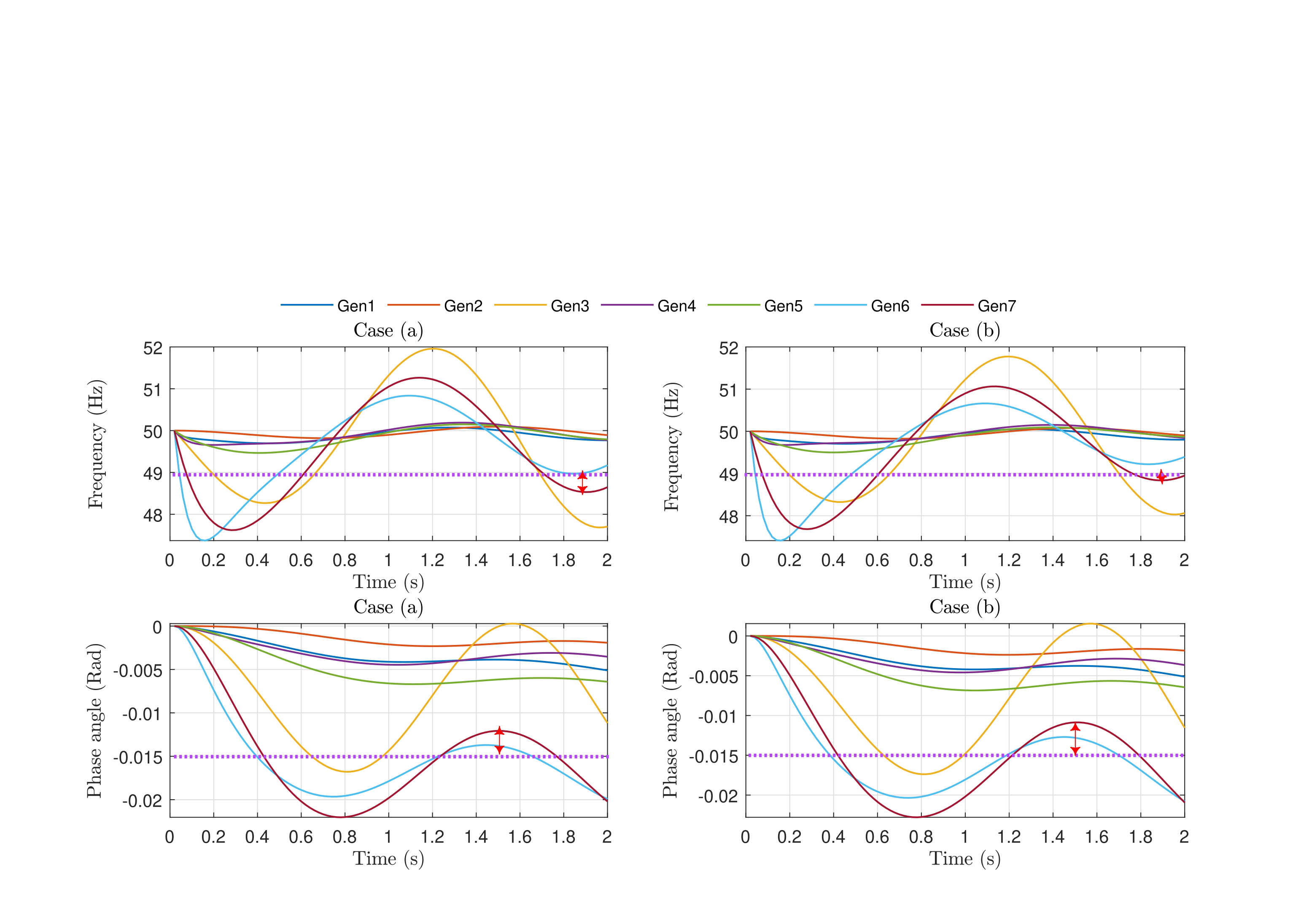}
  \caption{Comparing similar D-LAAs in IEEE 57-bus system: Case (a): $K_L(7,3) = 46.65$pu and $\epsilon_{6} =2.2$MW, , Case (b): $K_L(39,3) = 52.15$pu and $\epsilon_{6} =2.2$ MW}\label{fig:similar cases}
\end{figure*}

This section compares the general performance of the proposed approach to that of other benchmarks in common practical settings, which is shown in Table~\ref{tab:my-table_localization}. In smaller cases, such as the IEEE 14-bus system, all the algorithms achieve acceptable performance, particularly for single-point attacks; we also observe an approximate reduction of 1\% accuracy for multi-point attacks. The larger cases, such as the IEEE 39- and 57-bus systems, present more challenging scenarios. \textcolor{black}{In IEEE 39-bus and 57-bus systems, the performance gap between CN and the second-best approach (2D-CNN) is 3\% and 6\%, respectively.} In IEEE 57-bus, particularly for multi-point attacks, MLP and SVM fall below 80\% accuracy, and CNN-based techniques (i.e., 1D-CNN and 2D-CNN) cannot exceed 90\% accuracy, while CN achieves 94\% accuracy, demonstrating the importance of the vector representation of neurons and dynamic routing arrangements in CN. To visually examine the difference between CN and 2D-CNN (that achieves the closest performance to CN) in the localization of D-LAAs, Fig.\ref{fig:similar cases} depicts two similar attack cases in the IEEE 57-bus system (the most challenging scenario), Case (a): $K_L(7,3) = 46.65$pu, $\epsilon_6 = 2.2$MW and Case (b): $K_L(39,3) = 52.15$pu, $\epsilon_6 = 2.2$MW. Comparing the frequency and phase angle profiles of Cases (a) and (b) around the frequency of 49Hz and -0.015 Rad (dashed lines) reveals the most discernible difference between these two profiles, particularly the positions of the observed features. As stated in Section~\ref{why capsule nets}, CN is able to record the position of the detected entity or entity portions as capsule dimensions. By applying max-pooling layers instead of dynamic routing agreements in feature extraction (dimension reduction in feedforward flow of features), CNN-based techniques erase the positional information of observed features. This explains why CNs can correctly differentiate these circumstances and pinpoint the location of the attack, while 2D-CNN cannot. 



\begin{table}
\centering
\caption{Localization results for different IEEE cases}
\label{tab:my-table_localization}
\scalebox{0.70}{%
\begin{tabular}{ccccccc}
\hline
Method & \multicolumn{2}{c}{IEEE 14-bus} & \multicolumn{2}{c}{IEEE 39-bus} & \multicolumn{2}{c}{IEEE 57-bus} \\ \hline
 &
  \begin{tabular}[c]{@{}c@{}}Single-point\\ attack\end{tabular} &
  \begin{tabular}[c]{@{}c@{}}Multi-point\\ attack\end{tabular} &
  \begin{tabular}[c]{@{}c@{}}Single-point\\ attack\end{tabular} &
  \begin{tabular}[c]{@{}c@{}}Multi-point\\ attack\end{tabular} &
  \begin{tabular}[c]{@{}c@{}}Single-point\\ attack\end{tabular} &
  \begin{tabular}[c]{@{}c@{}}Multi-point\\ attack\end{tabular} \\ \hline
CN     & 99.07\%               & 98.29\%               & 98.32\%              & 97.01\%              & 95.50\%              & 94.00\%              \\ \hline
2D-CNN & 99.02\%              & 98.07\%              & 95.53\%              & 93.57\%              & 89.72\%              & 88.09\%              \\ \hline
1D-CNN & 97.17\%              & 96.55\%              & 89.76\%              & 87.59\%              & 84.34\%              & 81.66\%              \\ \hline
MLP    & 96.44\%              & 96.29\%              & 85.21\%              & 81.23\%              & 77.19\%              & 70.92\%              \\ \hline
SVM    & 95.13\%              & 89.36\%              & 70.15\%              & 69.93\%              & 62.84\%             & 57.27\%              \\ \hline
\end{tabular}%
}
\end{table}



\subsubsection{Impact of the noisy PMU data}\label{sec:numerical_simulation_noise} The existence of noise in input data is a challenge for data-driven methodologies in real-world applications. In this instance, the PMU data noise could well be approximated by a Gaussian distribution \cite{noise_li2021power}. Tables \ref{tab:my-table_localization_noise_single} and \ref{tab:my-table_localization_noise_multi} detail the performance of the proposed approach and other benchmarks in a variety of noisy situations ($26$, $20$, and $16.5$ dB SNR ($20 \log (signal/noise)$\cite{brown2016characterizing}) for single-point and multi-point D-LAAs, respectively. 
 CNs and 2D-CNNs approaches that read data in 2D-manner are more resistant to environmental noise in both single-point and multi-point attacks; for instance, in multi-point attacks on the IEEE 57-bus system, other benchmarks (i.e. 1D-CNN, MLP, and SVM) attain less than 80\% accuracy. Further investigation reveals that in the IEEE 57-bus case, there is roughly 7\% and 10\% difference in the accuracy of the localization task between the CN and 2D-CNN for single-point and multi-point attacks, respectively. Once again, these results show the superiority of CNs in localizing attacks in a noisy environment.

\begin{table}
\centering
\caption{Localization results for single-point attacks with noise}
\label{tab:my-table_localization_noise_single}
\scalebox{0.68}{%
\begin{tabular}{cccccccccc}
\hline
Method & \multicolumn{3}{c}{IEEE 14-bus} & \multicolumn{3}{c}{IEEE 39-bus} & \multicolumn{3}{c}{IEEE 57-bus} \\ \hline
 &
  \begin{tabular}[c]{@{}c@{}}26 DB\\ SNR\end{tabular} &
  \begin{tabular}[c]{@{}c@{}}20 DB\\ SNR\end{tabular} &
  \begin{tabular}[c]{@{}c@{}}16.5DB\\ SNR\end{tabular} &
  \begin{tabular}[c]{@{}c@{}}26 DB\\ SNR\end{tabular} &
  \begin{tabular}[c]{@{}c@{}}20 DB\\ SNR\end{tabular} &
  \begin{tabular}[c]{@{}c@{}}16.5DB\\ SNR\end{tabular} &
  \begin{tabular}[c]{@{}c@{}}26 DB\\ SNR\end{tabular} &
  \begin{tabular}[c]{@{}c@{}}20 DB\\ SNR\end{tabular} &
  \begin{tabular}[c]{@{}c@{}}16.5DB\\ SNR\end{tabular} \\ \hline
CN     & 97.89\%          & 97.45\%         & 95.12\%         & 97.19\%         & 96.71\%        & 91.69\%        & 94.89\%         & 94.11\%        & 90.41\%        \\ \hline
2D-CNN & 97.01\%         & 96.07\%        & 90.92\%        & 95.04\%         & 92.28\%        & 88.76\%        & 88.17\%         & 86.15\%        & 83.91\%        \\ \hline
1D-CNN & 96.74\%         & 96.08\%        & 89.13\%        & 87.93\%         & 86.15\%        & 83.59\%        & 83.87\%         & 80.02\%        & 77.92\%        \\ \hline
MLP    & 96.41\%         & 95.13\%        & 88.71\%        & 87.33\%         & 82.33\%        & 80.66\%        & 75.66\%         & 74.87\%        & 69.91\%        \\ \hline
SVM    & 91.07\%         & 90.39\%        & 87.84\%        & 62.91\%         & 60.44\%        & 54.79\%        & 64.12\%         & 54.82\%        & 52.41\%       \\ \hline
\end{tabular}%
}
\end{table}

\begin{table}
\centering
\caption{Localization results for multi-point attacks with noise}
\label{tab:my-table_localization_noise_multi}
\scalebox{0.68}{%
\begin{tabular}{cccccccccc}
\hline
Method & \multicolumn{3}{c}{IEEE 14-bus} & \multicolumn{3}{c}{IEEE 39-bus} & \multicolumn{3}{c}{IEEE 57-bus} \\ \hline
 &
  \begin{tabular}[c]{@{}c@{}}26 DB\\ SNR\end{tabular} &
  \begin{tabular}[c]{@{}c@{}}20 DB\\ SNR\end{tabular} &
  \begin{tabular}[c]{@{}c@{}}16.5DB\\ SNR\end{tabular} &
  \begin{tabular}[c]{@{}c@{}}26 DB\\ SNR\end{tabular} &
  \begin{tabular}[c]{@{}c@{}}20 DB\\ SNR\end{tabular} &
  \begin{tabular}[c]{@{}c@{}}16.5DB\\ SNR\end{tabular} &
  \begin{tabular}[c]{@{}c@{}}26 DB\\ SNR\end{tabular} &
  \begin{tabular}[c]{@{}c@{}}20 DB\\ SNR\end{tabular} &
  \begin{tabular}[c]{@{}c@{}}16.5DB\\ SNR\end{tabular} \\ \hline
CN     & 97.51\%         & 94.11\%         & 93.67\%         & 95.53\%         & 94.79\%        & 90.12\%        & 93.28\%         & 90.62\%        & 89.31\%        \\ \hline
2D-CNN & 96.14\%         & 93.13\%        & 88.32\%        & 92.51\%         & 86.63\%        & 88.32\%        & 84.33\%         & 81.89\%        & 79.31\%        \\ \hline
1D-CNN & 95.11\%         & 90.23\%        & 85.98\%        & 86.56\%         & 84.95\%        & 85.98\%        & 79.91\%         & 77.62\%        & 75.71\%        \\ \hline
MLP    & 96.66\%         & 90.61\%        & 81.31\%        & 85.98\%         & 79.43\%        & 81.31\%        & 67.33\%         & 66.19\%        & 59.76\%        \\ \hline
SVM    & 89.31\%         & 85.43\%        & 83.97\%        & 81.13\%         & 58.12\%        & 53.97\%        & 57.13\%         & 56.22\%        & 52.32\%        \\ \hline
\end{tabular}%
}
\end{table}

\subsubsection{Impact of the missed/outlier PMU data}\label{sec:numerical_simulation_missed/outlier}
\begin{figure}[htb]
    \centering
  \includegraphics[height=7.9 cm,width=8.9 cm,trim= 15 10 10 151,clip]{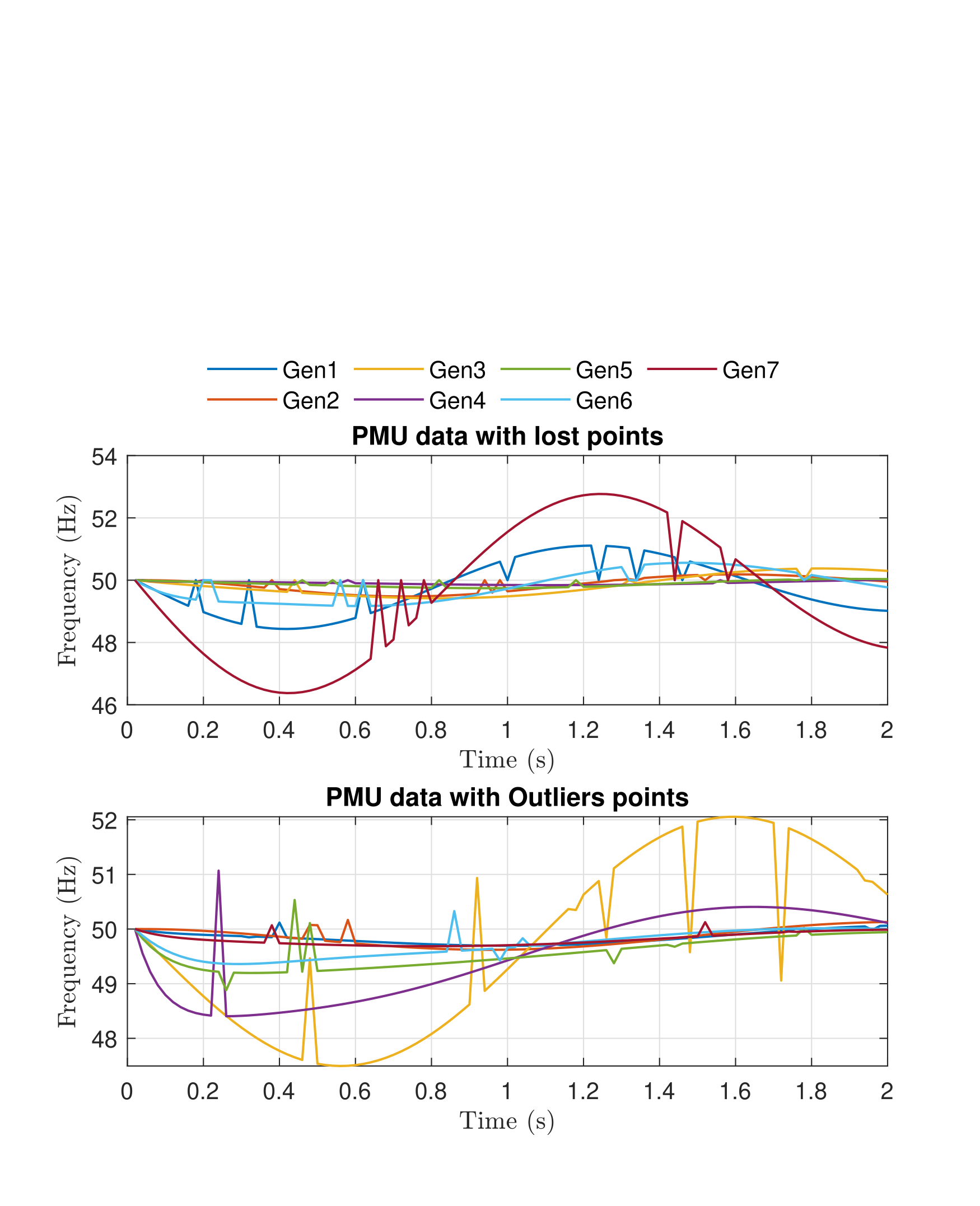}
  \caption{Frequency profiles for IEEE 57-bus system with lost points in PMU data (multi-point attack: $K_L(9,7)=43.31 pu$, $\epsilon_{35}= 0.14 $MW, $\epsilon_{29}= 0.43$MW) and outlier points in PMU data (multi-point attack: $K_L(11,3)$=$18.25pu$, $\epsilon_{3}=1.16$MW, $\epsilon_{29}=0.39$MW).}\label{fig:lost_outliers_data}
\end{figure}

Another point of concern in real applications is the presence of missing or outlier points in the input data. Two instances were explored in the simulation results to investigate these circumstances: (i) between 3\% and 5\% of randomly selected data points are removed and replaced with fixed reference values of 50 Hz and 0 Rad for frequency and phase angle profiles, respectively. (ii) 3\% to 8\% of the randomly selected points are replaced with new points that vary randomly within a 20\% range of the primary value. Fig. \ref{fig:lost_outliers_data} depicts samples of the previously described missing and outlier points. PMU data with outlier points, as shown in Fig. \ref{fig:lost_outliers_data}, provide a more challenging circumstance than PMU data with lost points owing to spikes. The localization outcomes for various IEEE cases under multi-point attacks, as shown in Table \ref{tab:my-table_localization_lost and outlier points}, support this hypothesis. CNs outperform other techniques, notably in IEEE 39- and 57-bus cases; 2D-CNNs achieve the next best performance. 


\begin{table}
\centering
\caption{Localization results with lost and outlier points}
\label{tab:my-table_localization_lost and outlier points}
\scalebox{0.66}{%
\begin{tabular}{ccccccc}
\hline
Method & \multicolumn{2}{c}{IEEE 14-bus} & \multicolumn{2}{c}{IEEE 39-bus} & \multicolumn{2}{c}{IEEE 57-bus} \\ \hline
 &
  \begin{tabular}[c]{@{}c@{}}Lost points\\ in PMU data\end{tabular} &
  \begin{tabular}[c]{@{}c@{}}Outlier points\\ in PMU data\end{tabular} &
  \begin{tabular}[c]{@{}c@{}}Lost points\\ in PMU data\end{tabular} &
  \begin{tabular}[c]{@{}c@{}}Outlier points\\ in PMU data\end{tabular} &
  \begin{tabular}[c]{@{}c@{}}Lost points\\ in PMU data\end{tabular} &
  \begin{tabular}[c]{@{}c@{}}Outlier points\\ in PMU data\end{tabular} \\ \hline
CN     & 94.11\%              & 91.29\%              & 90.33\%              & 88.49\%              & 88.42\%              & 85.09\%              \\ \hline
2D-CNN & 91.39\%              & 89.12\%              & 87.06\%              & 83.58\%              & 84.13\%              & 80.79\%              \\ \hline
1D-CNN & 89.23\%              & 83.82\%              & 84.51\%              & 78.63\%              & 77.29\%              & 71.11\%              \\ \hline
MLP    & 78.67\%              & 73.77\%              & 74.63\%              & 68.09\%              & 65.55\%              & 62.55\%              \\ \hline
SVM    & 49.13\%              & 46.97\%              & 43.19\%              & 35.47\%              & 39.33\%              & 34.81\%             \\ \hline
\end{tabular}%
}
\end{table}

\subsubsection{Impact of Detection Delays on Localization Performance}\label{sec:numerical_simulation_delayed}
\begin{figure}
    \centering
  \includegraphics[height=8.8 cm,width=9 cm,trim= 15 20 30 165,clip]{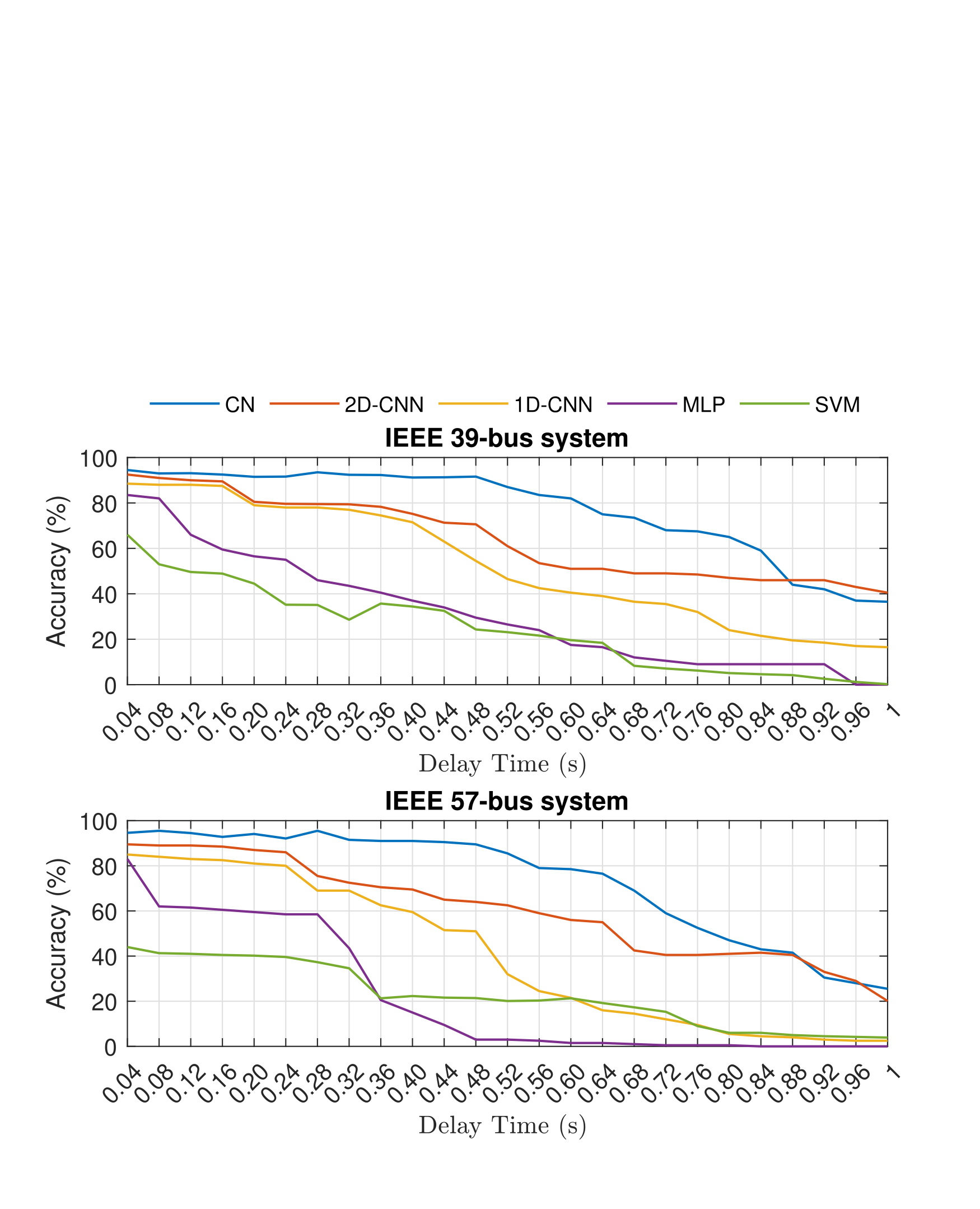}
  \caption{Impact of delay in PMU data on localization outcomes in IEEE 57-bus case}\label{fig:PMU delay}
\end{figure}

Finally, we examine the resilience of the proposed CN algorithm against detection delays {(as explained in Section \ref{introduction})}, which results in an offset between the training and testing datasets (as illustrated in  Fig. \ref{fig:delay_effect}). In particular, during the testing phase, we input data sets with varying delay times -- ranging from 0 to 1-second intervals in steps of 0.1-second. Fig. \ref{fig:PMU delay} shows a thorough comparison of the various approaches in the IEEE 39- and 57-bus cases. In comparison to existing benchmarks, CN performs well against delay in PMU data, losing 80\% accuracy after 0.60 seconds, whereas CNN-based techniques (2D-CNN and 1D-CNN) lose 80\% accuracy after 0.24 seconds. The prominent characteristic of the CN that maintains the instantiation parameters in a vector structure (i.e., 16- and 32-dimensional capsules for IEEE 39- and 57-bus systems) can tremendously aid us in these circumstances. The dynamic routing agreement between the Primary and Digit capsules empowers CN to make the most of these multi-dimensional features. On the other hand, CNN-based approaches eliminate the positional instantiations that are essential in these circumstances. Finally, NN and SVM, which read data point by point, entirely collapse due to data delay (after a 0.3-second delay, accuracy falls below 40\%.), which is the fundamental issue with these methods in practical applications.

\section{ Conclusions }


This paper presents a data-driven method based on CNs for monitoring and protecting the power grid from attacks against high-wattage IoT-enabled devices. We designed a CN architecture for power grid security monitoring using PMU data. CN improves the localization of impending D-LAAs by delivering capsule-based vector feature extraction and routing agreement. To validate CN's success in localization, numerical results were obtained on IEEE 14-, 39-, and 57-bus systems with single- and multi-point attacks, and CN offered between 94\% and 99\% accuracy. To examine the robustness of the suggested method in practice, CN and other benchmark techniques were tested using noisy, missing, outlier points, and delayed input data. CN excels in all of the above situations (delivering at least a 5\% improvement in precision over the best method afterward). Based on the numerical findings of this study in real-world circumstances, the robust feature extraction provided by the CN structure is recommended for securing power grids against cyberattacks.

\section*{Appendix: Simulation Parameters}
\small
{

Dynamic parameters for IEEE 14-bus system:

\begin{align*}
    & M_1 - M_5 = [0.125; 0.034; 0.016; 0.010; 0.015]; \\
    & D_1 - D_5 = [0.125; 0.068; 0.032; 0.068; 0.072]; \\
    & K^P_1 - K^P_{5}= [0.02; 0.09; 0.03; 0.03; 0.08]; \\
    & K^I_1 - K^I_5  =[0.35; 0.40; 0.35; 0.35; 0.40];\\
    &D_i = 0.01, \forall i \in \mathcal{N}_L;
\end{align*}}

Dynamic parameters for IEEE 39-bus system:
\begin{align*}
   & M_1 = 2.3186; 
   M_2 : M_8 = 2.6419;
    M_9 : M_{10}  = 2.4862. \\
    & K^P_1 - K^P_{10}= [1; 0.45; 0.45; 0.1; 0.5; 0.4; 0.3; 0.2; 0.4; 0.5]; \\
    & K^I_i = 0.6,  \forall i \in \mathcal{N}_G; D_i = 2,  \forall i \in \mathcal{N}_G;\\ 
    &D_i = 0.01, \forall i \in \mathcal{N}_L;
\end{align*}

Dynamic parameters for IEEE 57-bus system:
\begin{align*}
    & M_1 - M_7 = [2.6309;1.200;5.078;1.200;2.6309;1.200;2.6309];\\
    & D_1 - D_7 = [2;0;2;0;2;0;2];\\
    & K^P_1 - K^P_{7}= [25; 35; 10; 20; 30; 10; 30]; \\
    & K^P_1 - K^P_{7}= [25; 20; 20; 20; 30; 15; 30];\\
    & D_i = 0.2, \forall i \in \mathcal{N}_L;
\end{align*}

\bibliographystyle{IEEEtran}
\bibliography{ref}

\newpage

\end{document}